\documentclass[twocolumn]{article}

\usepackage{booktabs} % For formal tables

\usepackage{color}
\usepackage{booktabs}
\usepackage{tabulary}
\usepackage{siunitx}
\usepackage{tikz}
\usepackage{graphicx}
\usepackage{latexsym}
\usepackage{subfigure}
\usepackage{multirow}
\usepackage[xindy]{glossaries}
\usepackage{verbatim}
\usepackage{moreverb}
\usepackage{graphicx}
\usepackage{listings}
\usepackage{float}
\usepackage{graphics}
\usepackage{paralist}
\usepackage{amssymb}
\usepackage{amsmath}
\usepackage{listings}
\usepackage{setspace}
\usepackage[T1]{fontenc}

\usepackage[utf8]{inputenc}
\usepackage{newunicodechar} 
\usepackage{lstautogobble}
\usepackage{times}
\usepackage{latexsym}
\usepackage{lipsum}
\usepackage{calligra}
\usepackage{enumerate}
\usetikzlibrary{arrows,automata, shapes.symbols,patterns}

\newtheorem{definition}{Definition}

\usepackage{tgpagella}
\usepackage{textcomp}
\usepackage{authblk}
\usepackage{url}

\providecommand{\keywords}[1]{\textbf{\textit{Index terms---}} #1}
\begin{document}

\title{A Road-map Towards Explainable Question Answering \\ A Solution for Information Pollution}

\author[1]{Saeedeh Shekarpour}
\author[2]{Faisal Alshargi}

\affil[1]{Email: sshekarpour1@udayton.edu, University of Dayton, Dayton, USA}
\affil[2]{Email: alshargi@informatik.uni-leipzig.de, University of Leipzig, Leipzig, Germany}

\maketitle

\begin{abstract}
The increasing rate of information pollution on the Web requires novel
solutions to tackle that. Question Answering (QA) interfaces are simplified and user-friendly interfaces to access information on the Web. However, similar to other AI applications, they are black boxes which do not manifest the details of the learning or reasoning steps for augmenting an answer.
The Explainable Question Answering (XQA)  system can alleviate the pain of information pollution where it provides transparency to the underlying computational model and exposes an interface enabling the end-user to access and validate provenance, validity, context, circulation, interpretation, and feedbacks of information.
This position paper sheds light on the core concepts, expectations, and challenges in favor of the following questions (i) What is an XQA system?, (ii) Why do we need XQA?, (iii) When do we need XQA? (iv) How to represent the explanations? (iv) How to evaluate XQA systems?
\end{abstract}

%%
%% The code below is generated by the tool at http://dl.acm.org/ccs.cfm.
%% Please copy and paste the code instead of the example below.
%%

%%
%% Keywords. The author(s) should pick words that accurately describe
%% the work being presented. Separate the keywords with commas.

\section{Introduction}
Question Answering (QA) applications are a subcategory of Artificial Intelligence (AI) applications where for a given question, an adequate answer(s) is provided to the end-user regardless of concerns related to the structure and semantics of the underlying data.
The spectrum of QA implementations varies from statistical approaches \cite{shekarpour:www2013,sina}, deep learning models \cite{xiong2016dynamic,shekarpour:www2013} to simple rule-based (i.e., template-based) approaches \cite{www/UngerBLNGC12,ShekarpourANGHS11}. 
Also, the underlying data sets in which the answer is exploited might range from Knowledge Graphs (KG) holding a solid semantics as well as structure to unstructured corpora (free text) or consolidation of both.
Apart from the implementation details and the background data, roughly speaking, the research community introduced the following categories of QA systems:
\vspace{-0.2cm}
\begin{itemize}
   \item \textbf{Ad-hoc QA:} advocates simple and short questions and typically relies on one single KG or Corpus.
    \item \textbf{Hybrid QA:} requires federating knowledge from heterogeneous sources \cite{bast2007ester}. 
    \item \textbf{Complex QA:} deals with complex questions which are long, and ambiguous. Typically, to answer such questions, it is  required to exploit answers from a hybrid of KGs and textual content \cite{CQA/Somaye}.
    \item \textbf{Visualized QA:} answers texual questions from images \cite{emnlp/LiFYML18}.
    \item \textbf{Pipeline-based QA:} provides automatic integration of the state-of-the-art QA implementations \cite{www/SinghRBSLUVKP0V18,esws/SinghBRS18}.
\end{itemize}

\begin{figure*}[hpt!]
\scriptsize
\centering
  \includegraphics[height=4cm,width=0.90\textwidth]{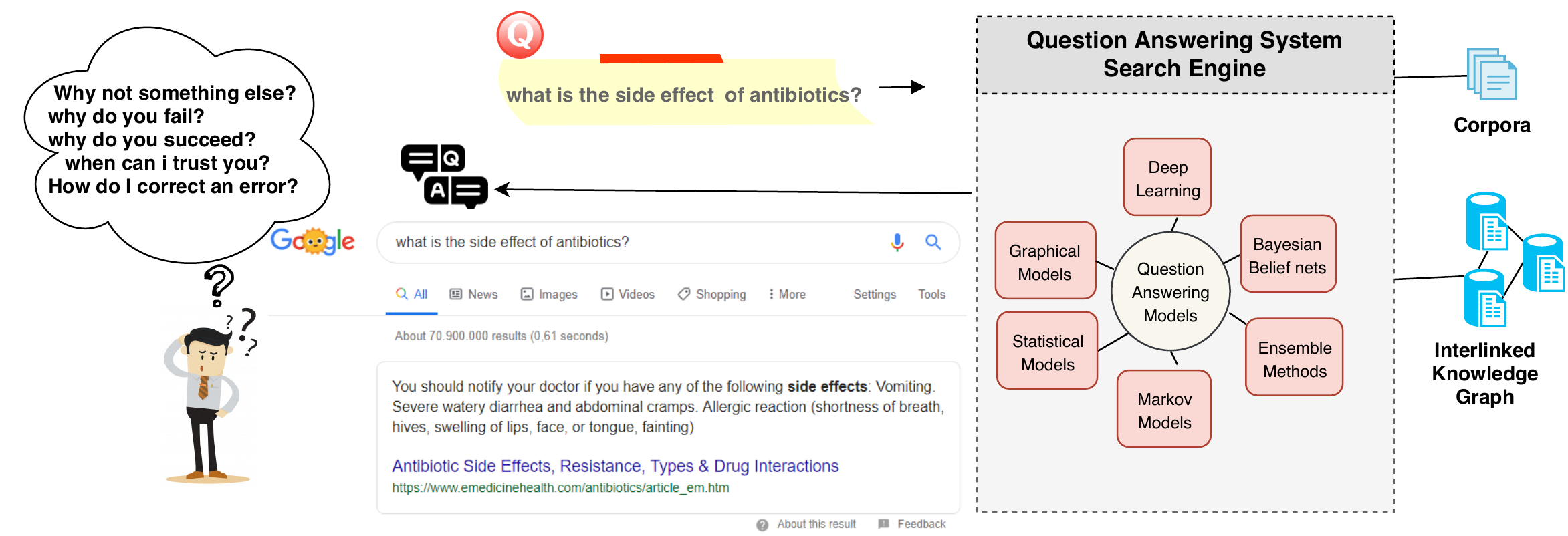}
  \caption{The existing QA systems are a black box which do not provide any explanation for their inference.}
  \label{fig:BlackBoxQA}
\end{figure*}

A missing point in all types of QA systems is that in case of either success or failure, they are silent to the question of why? Why have been a particular answer chosen? Why were the rest of the candidates disregarded? Why did the QA system fail to answer? whether it is the fault of the model, quality of data, or lack of data?
The truth is that the existing QA systems similar to other AI applications are a black box (see Figure \ref{fig:BlackBoxQA}) meaning they do not provide any supporting fact (explanation) about the represented answer with respect to the trustworthiness rate to the source of information, the confidence/reliability rate to the chosen answer, and the chain of reasoning or learning steps led to predict the final answer.
For example, Figure \ref{fig:BlackBoxQA} shows that the user sends the question \texttt{`what is the side effect of antibiotics?'} to the QA system.
If the answer is represented in a way similar to the interface of Google, then the end-user might have a mixed feeling as to whether s/he can rely on this answer or how and why such an answer is chosen among numerous candidates?

\begin{figure}[hp!]
    \centering
    \includegraphics[width=0.80\columnwidth]{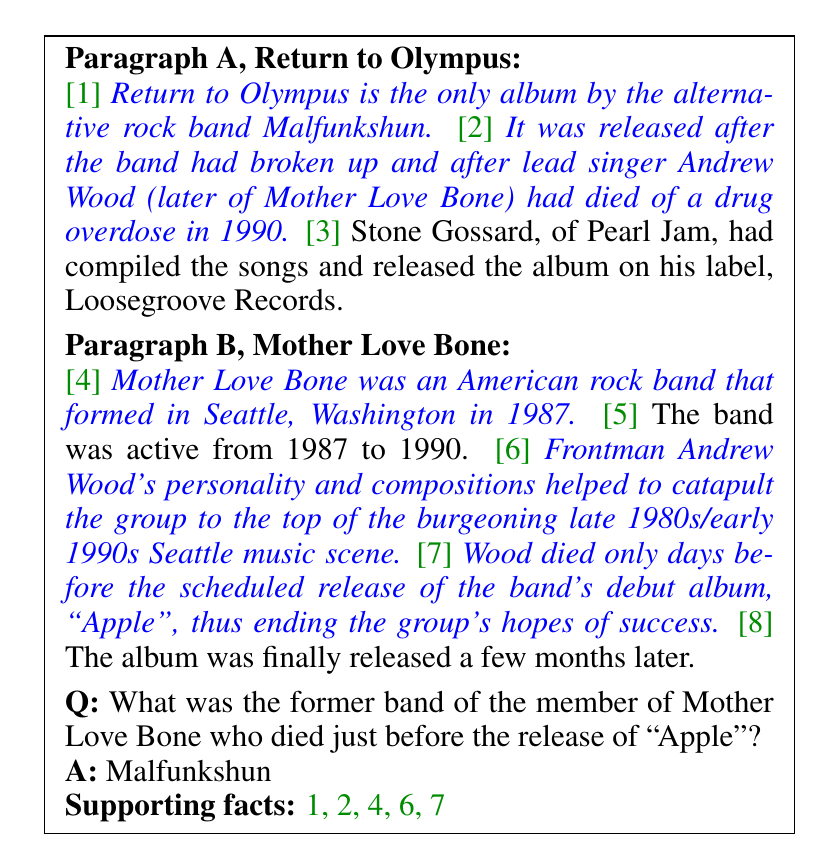}
    \caption{An example from \cite{HotpotQA} where the supporting facts necessary to answer the given question \textbf{Q} are listed. }
    \label{fig:HotPot-example}
\end{figure}

The rising challenges regarding the credibility, reliability, and validity of the state-of-the-art QA systems are of high importance, especially on critical domains such as life-science involved with human life.
The \textbf{Explainable Question Answering (XQA) systems} are an emerging area which tries to address the shortcomings of the existing QA systems.
The recent article \cite{HotpotQA} published a data set containing pairs of question/answer along with the supporting facts of the corpus where an inference mechanism over them led to the answer. 
Figure \ref{fig:HotPot-example} is an example taken from the original article \cite{HotpotQA}.
The assumption behind this data set is that the questions require multi-hops to conclude the answer, which is not the case all the time. 
Besides, this kind of representations might not be an ideal form for XQA; for example, whether representing solely the supporting facts is sufficient? how reliable are the supporting facts? Who published them? And how credible is the publisher? And furthermore, regarding the interface, is not the end-user overwhelmed if s/he wants to go through all the supporting facts? Is not there a more user-friendly approach for representation?

The XQA similar to all applications of Explainable AI (XAI) is expected to be transparent, accountable and fair \cite{XAI/guardian}. 
If QA is biased (bad QA), it will come up with discriminating information which is biased based on race, gender, age, ethnicity, religion, social or political rank of publisher and targeted user \cite{XAI/bad}. 
\cite{gunning2017explainable} raises six fundamental competency questions regarding XAI as follows:
\begin{enumerate}
    \item Why did the AI system do that?
    \item Why did not the AI system do something else?
    \item When did the AI system succeed?
    \item When did the AI system fail?
    \item When does the AI system give enough confidence in the decision that you can trust?
    \item How can the AI system correct an error?
    
\end{enumerate}

In the area of XQA, we adopt these questions; however, we apply sufficient modifications as follows:
\begin{enumerate}
    \item Why did the QA system choose this answer?
    \item Why did not the QA system answer something else?
    \item When did the QA system succeed?
    \item When did the QA system fail?
   \item  When does the QA system give enough confidence in the answer that you can trust?
    \item How can the QA system correct an error?
    
\end{enumerate}

This visionary paper introduces the core concepts, expectations and challenges in favor of the questions (i) What is an Explainable Question Answering (XQA) system?, (ii) Why do we need XQA?, (iii) When do we need XQA? (iv) How to represent the explanations? (iv) How to evaluate XQA systems?
In the following sections, we address each question respectively.

\begin{figure*}[hpbt!]
\scriptsize
\centering
  \includegraphics[height=6cm,width=0.9\textwidth]{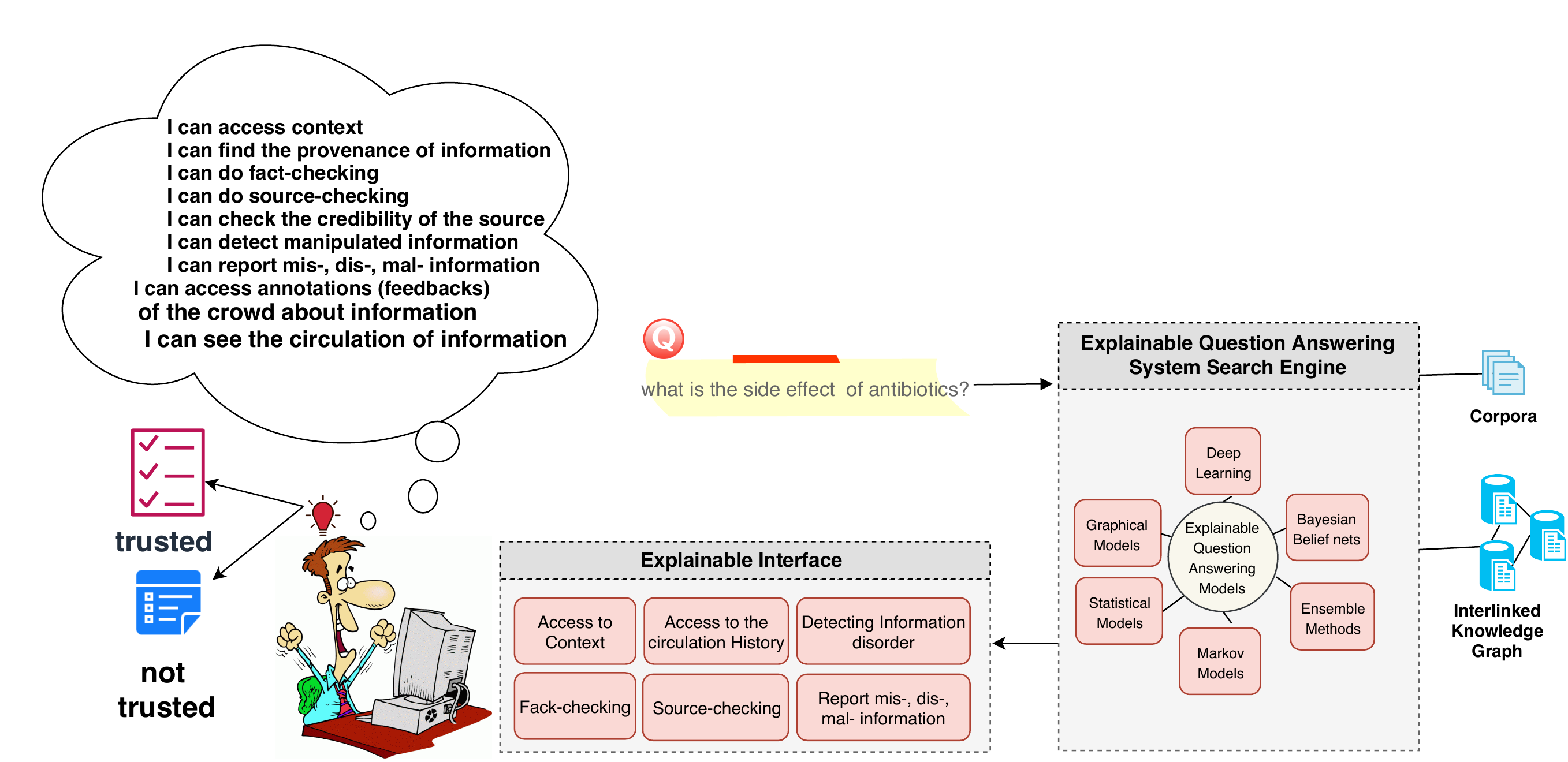}
  \caption{The explainable question answering exposes explainable models and explainable interface; then the user can make a decision as to whether to trust or not.}
  \label{fig:xqa}
\end{figure*}

\section{What is XQA?}

To answer the question of \textbf{What is XQA?}, we feature two layers i.e., \textbf{model} and \textbf{interface} for XQA similar to XAI \cite{gunning2017explainable}.
Figure \ref{fig:xqa} shows our envisioned plan for XQA where at the end, the end user confidently conclude that he can/cannot trust to the answer.
In the following, we present a formal definition of XQA.

\begin{definition}[Explainable Question Answering]
XQA is a system relying on an \textbf{explainable computational model} for exploiting the answer and then utilizes an \textbf{explainable interface} to represent the answer(s) along with the explanation(s) to the end-user.  
\end{definition}

This definition highlights two major components of XQA as (i) \textbf{explainable computational model} and (ii) \textbf{explainable interface}. In the following we discuss these two components in more details:

\paragraph{\textbf{Explainable Computational Model.}}
Whatever computational model employed in XQA system, (e.g., learning-based model, schema-driven approach, reasoning approach, heuristic approach, rule-based approach, or a mixture of various models) it has to explain all intermediate and final choices meaning the rationale behind the decisions should be \textbf{transparent}, \textbf{fair}, and \textbf{accountable} \cite{XAI/guardian}.
The responsible QA system distinguishes misinformation, disinformation, mal-information, and true facts \cite{informationdisorder}. 
Furthermore, it cares about the untrustworthiness and trustworthiness of data publisher, information representation, updated or outdated information, accurate or inaccurate information, and also the interpretations that the answer might raise.
Whereas, the fair QA system is not biased based on the certain characteristics of the data publisher, or the targeted end user (e.g., region, race, social or political rank).
Finally, the transparency of QA systems refers to the availability and accessibility to the reasons behind the decisions of the QA system in each step upon the request of involving individuals (e.g., end user, developer, data publisher, policymakers).

\paragraph{\textbf{Explainable Interface.}}
The explainable interface introduced in \cite{gunning2017explainable} contains two layers (i) a cognitive layer and (ii) an explanation layer.
The cognitive layer represents the implications learned from the computational model in an explainable form (abstractive or summarized representation), and then the explanation layer is responsible for delivering them to the end user in an interactive mode. 
We introduce several fundamental features which the future generation of XQA have to launch. We extensively elaborate on our view about the interface in Section 5.

\section{Why do we need XQA?}
We showcase the importance of having XQA using the two following arguments.

\paragraph{\textbf{Information Disorder Era.}}
The growth rate of mis-, dis-, mal- information on the Web is getting dramatically worsened \cite{informationdisorder2}.
Still, the existing search engines fail to identify misinformation even where it is highly crucial \cite{kata2010postmodern}.
It is expected from the information retrieval systems (either keyword-based search engines or QA systems) to identify mis-, dis-, mal- information from reliable and trustworthy information.

\paragraph{\textbf{Human Subject Area}.}
Having XQA for areas being subjected to lives particularly human subject is highly important.
For example, bio-medical and life-science domains require to discriminate between the hypothetical facts, resulting facts, methodological facts, or goal-oriented facts.
Thus XQA has to infer the answer of informational question based on the context of the question as to whether it is asking about resulting facts, or hypothetical facts, etc.

\section{When do we need XQA?} 
Typically in the domains that the user wants to make a decision upon the given answer, XQA matters since it enables the end user to make a decision with trust.
There are domains that traditional QA does not hurt. For example, if the end user is looking for the \texttt{`nearby Italian restaurant'}, QA systems suffice.
On the contrary, in the domain of health, having the explanations is demanding otherwise the health care providers can not entirely rely on the answers disposed by the system.

\section{How to represent the explanation?}
\begin{figure}[hpbt!]
\scriptsize
\centering
  \includegraphics[width=\columnwidth]{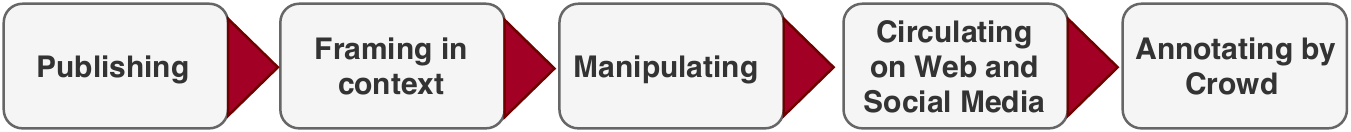}
  \caption{The life cycle of information on the Web.}
  \label{fig:lifecycle}
\end{figure}

Before, presenting the features of the interface; we illustrate the life cycle of information on the Web (see Figure \ref{fig:lifecycle}).
Each piece of information has a source who initially published it, then 
After reposting the original piece, it might be framed in a context or manipulated. Also, concerning its circulation on social media or the Web, it might be annotated or commented on by the crowd.

\begin{figure}[hpbt!]
\scriptsize
\centering
  \includegraphics[height=5cm, width=0.8\columnwidth]{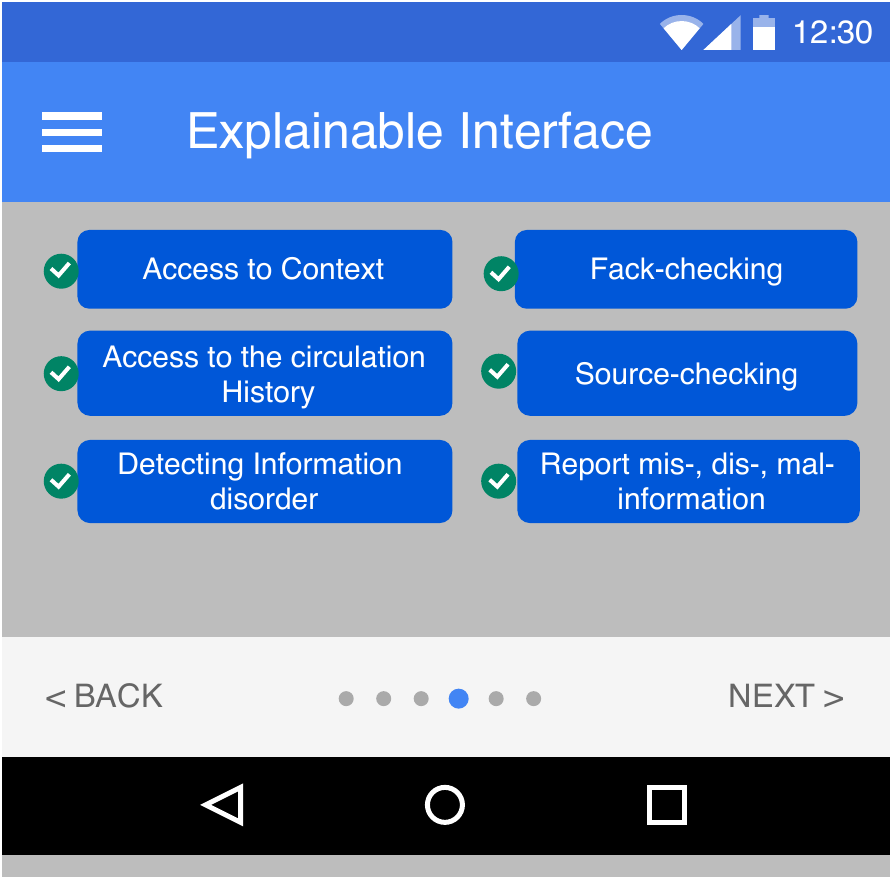}
  \caption{The features of explainable interface for XQA.}
  \label{fig:xinterface}
\end{figure}

We feature the explainable QA interface with respect to its life cycle in Figure \ref{fig:xinterface}. The explainable interface should enable the end user to
\begin{itemize}
    \item access context
    \item find the provenance of information
    \item do fact-checking
    \item do source-checking
    \item check the credibility of the source
    \item detect manipulated information
    \item report mis-, dis-, mal- information
    \item access annotations (feedbacks) of the crowd
    \item see the circulation history
\end{itemize}

\section{How to evaluate XQA systems?}
The evaluation of the XQA systems has to check the validity and performance of all the features in the computational model and interface. For example, whether the interface disposes a fact-checking feature or source checking feature? If yes, how fair, transparent, and accountable are these features?
The research community requires to introduce metrics, criteria, and benchmarks for evaluating these features.

\section{Conclusion}
In this paper, we discussed the concepts, expectations, and challenges of XQA.
The expectation is that the future generation of QA systems (or search engines) rely on computational explainable models and interact with the end-user via the explainable user interface.
The explainable computational models are transparent, fair and accountable.
Also, the explainable interfaces enable the end-user to interact with features for source-checking,  fact-checking and also accessing  to
 context and circulation history.
In addition, the explainable interfaces allow the end-user to report mis-, dis-, mal- information.

We are at the beginning of a long-term agenda to mature this vision and furthermore provide standards and solutions.
The phenomena of information pollution is a dark side of the Web which will endanger our society, democracy, justice service and health care.
We hope that the XQA will be the attention of the research community in the next couple of years.

%% The next two lines define the bibliography style to be used, and
%% the bibliography file.
\bibliographystyle{plain}
\bibliography{ref}

\begin{thebibliography}{10}

\bibitem{CQA/Somaye}
Somayeh Asadifar, Mohsen Kahani, and Saeedeh Shekarpour.
\newblock Hcqa: Hybrid and complex question answering on textual corpus and
  knowledge graph.
\newblock {\em CoRR}, abs/1811.10986, 2018.

\bibitem{bast2007ester}
Holger Bast, Alexandru Chitea, Fabian Suchanek, and Ingmar Weber.
\newblock Ester: efficient search on text, entities, and relations.
\newblock In {\em Proceedings of the 30th annual international ACM SIGIR
  conference on Research and development in information retrieval}, pages
  671--678. ACM, 2007.

\bibitem{XAI/bad}
Stephen Buranyi.
\newblock Rise of the racist robots – how ai is learning all our worst
  impulses.
\newblock
  \url{https://www.theguardian.com/inequality/2017/aug/08/rise-of-the-racist-robots-how-ai-is-learning-all-our-worst-impulses}.
\newblock Accessed: 2017-8-8.

\bibitem{gunning2017explainable}
David Gunning.
\newblock Explainable artificial intelligence (xai).
\newblock {\em Defense Advanced Research Projects Agency (DARPA), nd Web
  (2017)}.

\bibitem{kata2010postmodern}
Anna Kata.
\newblock A postmodern pandora's box: anti-vaccination misinformation on the
  internet.
\newblock {\em Vaccine}, 28(7):1709--1716, 2010.

\bibitem{emnlp/LiFYML18}
Qing Li, Jianlong Fu, Dongfei Yu, Tao Mei, and Jiebo Luo.
\newblock In {\em Proceedings of the 2018 Conference on Empirical Methods in
  Natural Language Processing, Brussels, Belgium, 2018}, pages 1338--1346,
  2018.

\bibitem{XAI/guardian}
Ian Sample.
\newblock Computer says no: why making ais fair, accountable and transparent is
  crucial.
\newblock
  \url{https://www.theguardian.com/science/2017/nov/05/computer-says-no-why-making-ais-fair-accountable-and-transparent-is-crucial}.
\newblock Accessed: 2017-11-05.

\bibitem{ShekarpourANGHS11}
Saeedeh Shekarpour, S{\"{o}}ren Auer, Axel{-}Cyrille~Ngonga Ngomo, Daniel
  Gerber, Sebastian Hellmann, and Claus Stadler.
\newblock Keyword-driven {SPARQL} query generation leveraging background
  knowledge.
\newblock In {\em Proceedings of the 2011 {IEEE/WIC/ACM} International
  Conference on Web Intelligence, {WI} 2011, Campus Scientifique de la Doua,
  Lyon, France, August 22-27, 2011}, pages 203--210, 2011.

\bibitem{sina}
Saeedeh Shekarpour, Edgard Marx, Axel{-}Cyrille~Ngonga Ngomo, and S{\"{o}}ren
  Auer.
\newblock {SINA:} semantic interpretation of user queries for question
  answering on interlinked data.
\newblock {\em J. Web Semant.}, 30:39--51, 2015.

\bibitem{shekarpour:www2013}
Saeedeh Shekarpour, Axel{-}Cyrille~Ngonga Ngomo, and S{\"{o}}ren Auer.
\newblock Question answering on interlinked data.
\newblock In {\em 22nd International World Wide Web Conference, {WWW} '13, Rio
  de Janeiro, Brazil, May 13-17, 2013}, pages 1145--1156, 2013.

\bibitem{esws/SinghBRS18}
Kuldeep Singh, Andreas Both, Arun~Sethupat Radhakrishna, and Saeedeh
  Shekarpour.
\newblock Frankenstein: {A} platform enabling reuse of question answering
  components.
\newblock In {\em The Semantic Web - 15th International Conference, {ESWC}
  2018, Heraklion, Crete, Greece, June 3-7, 2018, Proceedings}, pages 624--638,
  2018.

\bibitem{www/SinghRBSLUVKP0V18}
Kuldeep Singh, Arun~Sethupat Radhakrishna, Andreas Both, Saeedeh Shekarpour,
  Ioanna Lytra, Ricardo Usbeck, Akhilesh Vyas, Akmal Khikmatullaev, Dharmen
  Punjani, Christoph Lange, Maria{-}Esther Vidal, Jens Lehmann, and S{\"{o}}ren
  Auer.
\newblock Why reinvent the wheel: Let's build question answering systems
  together.
\newblock In {\em Proceedings of the 2018 World Wide Web Conference on World
  Wide Web, {WWW} 2018, Lyon, France, April 23-27, 2018}, pages 1247--1256,
  2018.

\bibitem{www/UngerBLNGC12}
Christina Unger, Lorenz B{\"{u}}hmann, Jens Lehmann, Axel{-}Cyrille~Ngonga
  Ngomo, Daniel Gerber, and Philipp Cimiano.
\newblock Template-based question answering over {RDF} data.
\newblock In {\em Proceedings of the 21st World Wide Web Conference 2012, {WWW}
  2012, Lyon, France, April 16-20, 2012}, pages 639--648, 2012.

\bibitem{informationdisorder2}
Claire Wardle.
\newblock Disinformation gets worse.
\newblock \url{https://www.niemanlab.org/2017/12/disinformation-gets-worse/}.
\newblock Accessed: 2018.

\bibitem{informationdisorder}
Claire Wardle and Hossein Derakhshan.
\newblock Information disorder: Toward an interdisciplinary framework for
  research and policymaking.
\newblock
  \url{https://shorensteincenter.org/information-disorder-framework-for-research-and-policymaking/}.
\newblock Accessed: 2017-9-31.

\bibitem{xiong2016dynamic}
Caiming Xiong, Stephen Merity, and Richard Socher.
\newblock Dynamic memory networks for visual and textual question answering.
\newblock In {\em International conference on machine learning}, pages
  2397--2406, 2016.

\bibitem{HotpotQA}
Zhilin Yang, Peng Qi, Saizheng Zhang, Yoshua Bengio, William~W. Cohen, Ruslan
  Salakhutdinov, and Christopher~D. Manning.
\newblock Hotpotqa: {A} dataset for diverse, explainable multi-hop question
  answering.
\newblock In {\em Proceedings of the 2018 Conference on Empirical Methods in
  Natural Language Processing, Brussels, Belgium, October 31 - November 4,
  2018}, pages 2369--2380, 2018.

\end{thebibliography}

%%
%% If your work has an appendix, this is the place to put it.

\end{document}